\documentclass[10pt,conference]{IEEEtran}
\IEEEoverridecommandlockouts
\usepackage{url}
\usepackage{cite}
\usepackage{array}
\usepackage{algorithmic}
\usepackage{graphicx}
\usepackage{textcomp}
\usepackage{xcolor}
\usepackage{tcolorbox}
\usepackage{longtable}
\usepackage{booktabs}
\usepackage{boldline}
\usepackage{makecell}
\usepackage{svg}
\usepackage[utf8]{inputenc}
\usepackage[absolute]{textpos}

\def\BibTeX{{\rm B\kern-.05em{\sc i\kern-.025em b}\kern-.08em
    T\kern-.1667em\lower.7ex\hbox{E}\kern-.125emX}}
\begin{document}

\begin{textblock*}{\paperwidth}(0pt,0pt)
    \vspace{2mm}
    \small
    \noindent \centering
    \begin{minipage}{0.7\paperwidth}
    \begin{tcolorbox}[left=0mm,right=0mm,boxrule=0.1mm,colback=white!30!white]
        \vspace{-2mm}
        \centering\emph{This is a PREPRINT version accepted to be published in the proceedings of the 31st International Conference on Software Analysis, Evolution, and Reengineering (SANER’24) - ERA Track. Users are expected to adhere to the terms and conditions set by each copyright holder when reproducing or making these documents available.}
        \vspace{-2mm}
    \end{tcolorbox}
    \end{minipage}
\end{textblock*}

\title{Navigating Expertise in Configurable Software Systems through the Maze of Variability}

\author{\IEEEauthorblockN{Karolina Milano}
\IEEEauthorblockA{\textit{Federal Institute of Mato Grosso do Sul} \\
Jardim, Brazil \\
karolina.milano@ifms.edu.br}
\and
\IEEEauthorblockN{Bruno Cafeo}
\IEEEauthorblockA{\textit{University of Campinas} \\
Campinas, Brazil \\
cafeo@unicamp.br}
}

\maketitle

\begin{abstract}
The understanding of source code in large-scale software systems poses a challenge for developers. The role of expertise in source code becomes critical for identifying developers accountable for substantial changes. However, in the context of configurable software systems (CSS) using pre-processing and conditional compilation, conventional expertise metrics may encounter limitations due to the non-alignment of variability implementation with the natural module structure. This early research study investigates the distribution of development efforts in CSS, specifically focusing on variable and mandatory code. It also examines the engagement of designated experts with variable code in their assigned files. The findings provide insights into task allocation dynamics and raise questions about the applicability of existing metrics, laying the groundwork for alternative approaches to assess developer expertise in handling variable code. This research aims to contribute to a comprehensive understanding of challenges within CSS, marking initial steps toward advancing the evaluation of expertise in this context.
\end{abstract}


\section{Introduction}

Software developers engage with numerous code elements, facing the challenge of comprehending an entire system's code. This complexity raises questions about determining the most suitable developer for maintenance tasks or identifying developers best equipped to review a pull request~\cite{steinmacher2015systematic }.

The concept of expertise in source code plays a crucial role in this context, guiding the identification of developers responsible for significant code changes. In large-scale software systems with diverse modules, expertise metrics are constantly used to coordinate developer teams, strategically distribute work, identify key collaborators, and select developers for specific tasks within a file~\cite{hannebauer2016automatically}. 

However, a significant challenge arises in the context of Configurable Software Systems (CSS). CSSs share a core with varied functionalities and configurations. The common core, so-called mandatory code, embodies fundamental functionality, while specific. In C-based CSS implementations, developers use the C preprocessor with \#ifdef directives and macro expressions to selectively include or exclude code segments based on predefined conditions, handling variabilities in the program (i.e., variable code).
In this case, variability implementation often diverges from the natural module structure (e.g. files), leading to fragmented and intertwined code associated with a single unit of variability~\cite{apel2013book,kastner2008granularity}. This misalignment could challenge code comprehension and maintenance, as conventional expertise metrics assume a module-based approach.

In this context, existing expertise metrics may not seamlessly apply to CSS, raising questions about task allocation, developer responsibilities, and the appropriateness of existing metrics such as: How many developers implement variabilities? Are they responsible for both mandatory and variable codes? Does a developer's expertise in a file correlate with expertise in variable code? 

This way, this early research study starts addressing these questions by seeking insights into task allocation among developers. Additionally, it aims to shed light on the applicability of expertise metrics in this context, potentially leading to the development of alternative approaches for assessing expertise in handling variable code. Our exploratory study was conducted across 25 open-source systems, encompassing a total of 450,255 commits carried out by 9,678 developers, and it is centered on the following research questions (RQs): RQ1 -  How are development efforts distributed among developers concerning variable and mandatory code? RQ2 - Do developers designated as experts actively engage with variable code in their designated files?

The results reveal the significant presence of developers primarily focused on mandatory code, potentially limiting their understanding of variable code. The implications underscore the importance of striking a balance in task allocation. Additionally, refining expertise metrics is crucial to ensure the accurate identification of developers with variable code knowledge. This, in turn, promotes a more equitable distribution of variable code tasks among developers. Additionally, the study explores precision and recall metrics for RQ2, demonstrating significant precision values. However, challenges in achieving high recall highlight the need for further refinement in expertise metrics to achieve a more comprehensive identification of developers experienced with variable code.

To the best of our knowledge, this is the first study that evaluates expertise metrics through the lens of variable code. Though further investigation is necessary, this study proves the importance of a study towards work specialization and its impacts on CSS.

\smallskip

\noindent\textbf{Replication Package.} Our code and data are shared at the following link: \url{https://doi.org/10.5281/zenodo.10176980}.

%The paper's structure is as follows: Section~\ref{sec:preliminaries} introduces basic CSS concepts and two expertise-related metrics. Our exploratory study is detailed in Section~\ref{sec:study}, with preliminary results presented in Section~\ref{sec:results}. Section~\ref{sec:implications} discusses the study's implications, while Section~\ref{sec:threats} addresses threats to validity. In Section~\ref{sec:relatedwork}, we review relevant literature, and Section~\ref{sec:conclusion} concludes the paper, highlighting directions for future work on the use of expertise metrics in CSS. 
\section{Background}
\label{sec:preliminaries}

To provide a contextual foundation for the forthcoming sections, we introduce concepts related to CSS and spotlight two widely used expertise metrics in the literature~\cite{avelino2019measuring, bird2011don}.

\subsection{Configurable Software Systems} 

%CSS are software systems that share a common core while also accommodating various functionalities and configurations~\cite{apel2013book}. The common core embodies the fundamental functionality present in all program variants within the family, while different configuration selections define the set of program variants. 

When examining CSS implemented in the C programming language (focus of this work), developers frequently employ the C preprocessor to annotate the implementation code for handling variabilities~\cite{apel2013book,kastner2008granularity}. The preprocessor identifies the code segments to be included or excluded from compilation based on preprocessor directives, particularly \texttt{\#ifdef} directives, in conjunction with macro expressions. These macro expressions can be constructed as boolean formulas comprising one or more macros, each referring to specific variabilities. Consequently, variabilities consist of program elements enclosed by preprocessor directives allowing their code to permeate throughout the program~\cite{kastner2008granularity}.

\subsection{Expertise Metrics}
Accurate assessment of developer expertise is crucial in collaborative software development. This section explores two key expertise metrics, namely Degree of Authorship (DOA) and Ownership. These metrics provide quantitative insights into developer contributions and responsibilities, facilitating informed decision-making and project management.

\medskip

\subsubsection{Degree of Authorship (DOA)} \label{sec:doa}

We employ the DOA metric proposed by Fritz et al.~\cite{fritz2014degree}. The DOA assessment considers three fundamental parameters. The first parameter involves determining whether the contributor initially created the file. The second parameter takes into account the frequency of changes made by the contributor in the file. Lastly, the third parameter reflects the count of modifications introduced by contributors other than the assessed contributor. These parameters collectively contribute to evaluating the degree of authorship for a given contributor in a specific file. For simplicity, we adopt the normalized DOA ($DOA_N$) proposed by Avelino~et~al., ranging from 0 to 1, with a threshold of 0.75 designating authors and contributors~\cite{avelino2019measuring}.

% To quantify developer expertise, we employ the \textit{degree-of-authorship (DOA)} metric proposed by Fritz et al.~\cite{fritz2014degree}. The DOA for contributor in a file and in a specific software version is defined as:

% \smallskip

% \begin{equation}
% \begin{aligned}
% DOA(\textit{c, f}) = 3.293 + 1.098 * FA + 0.164 * DL \\- 0.321 * \ln(1 + AC)
% \end{aligned}
% \end{equation}

% \smallskip

% The DOA of a contributor \textit{c} in a file \textit{f} is computed based on three primary parameters. First is the First Authorship (FA), a binary parameter indicating whether \textit{c} created file \textit{f} (1 if true, 0 otherwise). Second is the number of deliveries (DL), representing the count of changes made by \textit{c} in file \textit{f}. The third parameter is the number of acceptances (AC), corresponding to changes made by contributors other than \textit{c}. For simplicity, we adopt the normalized DOA ($DOA_N$) proposed by Avelino~et~al., ranging from 0 to 1, with a threshold of 0.75 designating authors and contributors~\cite{avelino2019measuring}.

\smallskip

\subsubsection{Ownership}

Ownership, an expertise metric for determining individual responsibility for a part of the source code, is calculated by examining the ratio of a developer's commits to the total commits for a specific file~\cite{bird2011don}. Interpretation relies on thresholds, with our study following the guidelines set by Bird~et~al.~\cite{bird2011don}. Developers with ownership below 5\% are categorized as minor contributors, while those with equal to or exceeding 5\% ownership are considered major contributors.
\smallskip
% \begin{equation}
% \begin{aligned}
% ratio = (developers commits x 100 )/Total commits in file
% \end{aligned}
% \end{equation}
\smallskip
\section{Exploratory Study}\label{sec:study}

Communities using CSS encounter a noticeable lack of guidance regarding effective code expertise management\\ within developer teams~\cite{Farooqui2019,Velez2022}. This study advocates for the establishment of an empirical knowledge base to explore expertise metrics within CSS, addressing this guidance deficit. We investigate the evolutionary histories of 25 systems, aiming to understand how variable code undergoes changes in the context of CSS over time. Simultaneously, we assess the applicability of established file-based expertise metrics, such as Degree of Authorship~\cite{fritz2014degree} and Code Ownership~\cite{bird2011don}, in capturing the nuances of variable code changes.

%It is crucial to acknowledge that the chosen expertise metrics, though not exhaustive, encompass fundamental concepts widely adopted within the field~\cite{rahman2011ownership,bird2011don,avelino2019measuring, cury2022}

\subsection{Research Questions} \label{sec:rq}

To accomplish the goal of this study, we address two fundamental research questions:

\medskip

\noindent\textit{RQ1: How are development efforts distributed among developers concerning variable and mandatory code?}

\smallskip

\noindent\underline{Rationale:} In a collaborative development environment, understanding how developers distribute their expertise across variable and mandatory code is essential. CSS, where variable code intertwines with mandatory code, poses a unique challenge. This investigation aims to reveal the distribution of development efforts, shedding light on whether developers engage with both code types or specialize in one of the types. By discerning this distribution, we seek insights into how developers navigate variable and mandatory code intricacies in CSS.

\medskip

\noindent\textit{RQ2: Do developers designated as experts actively engage with variable code in their designated files?}

\smallskip

\noindent\underline{Rationale:} In CSS, expertise metrics often designate developers as experts in specific files. It is crucial to ascertain whether recognized experts actively contribute to variable code within these files. This investigation ensures alignment between recommendations of expertise metrics and developers' practical experience, affirming the accuracy of conventional expertise metrics in CSS. A negative response suggests a potential misalignment, emphasizing the need for precise metrics in such a context.

\subsection{Subject Systems} \label{sec:systems}

This study analyzes expertise metrics within 25 real-world, open-source, CSS implemented in C. These systems, publicly hosted on GitHub, have undergone extensive examination in prior academic studies~\cite{abal2018variability, liebig2010analysis, medeiros2017discipline}. To refine our analysis, we selected systems with more than 30 developers. 

Table~\ref{tab:project-metrics} presents the analyzed systems and information about the number of files (\# Files), number of variabilities (\# Var.), number of commits (\# Commits), and number of developers (\# Devs.).

% \begin{figure}[htb] 
%     \centering
%     \includegraphics[width=\linewidth]{Systems.png} 
%     \caption{Work specialization among developers.}
%     \label{fig:Systems} 
% \end{figure}

% \begin{figure}[h]
%   \centering
%   \includesvg[width=0.35\linewidth]{Files.svg}
%   \includesvg[width=0.35\linewidth]{Var.svg}\\[1em]
%   \includesvg[width=0.45\textwidth]{Commits.svg}\hfill
%   \includesvg[width=0.45\textwidth]{Devs.svg}
%   \caption{Quatro imagens SVG organizadas em um layout 2x2.}
%   \label{fig:quatro-imagens}
% \end{figure}

\begin{table*}[t]
    \scriptsize
    \caption{Subject systems and their results.}
    \label{tab:project-metrics}
    \centering
    \begin{tabular}{lrrrr|c|rcrc}
        \toprule
        \textbf{Project} & \textbf{\# Files} & \textbf{\# Var.} & \textbf{\# Commits} & \textbf{\# Devs.} & \makecell{\textbf{Generalist / Specialist /} \\ \textbf{Mixed (\%)}} & \makecell{\textbf{DOA} \\ \textbf{ (\% of devs)}} & \makecell{\textbf{Precision / Recall} \\ \textbf{(DOA)}} & \makecell{\textbf{Ownership} \\ \textbf{(\%  of devs)}} & \makecell{\textbf{Precision / Recall} \\ \textbf{(Ownership)}} \\
        \midrule
        busybox & 1406 & 3344 & 14844 & 378 & 73.5 / 1.59 / 24.87 & 19.05 & 0.40 / 0.35 & 33.60 & 0.35 / 0.52 \\
        collectd & 645 & 847 & 6708 & 486 & 74.3 / 2.05 / 23.61 & 25.93 & 0.33 / 0.38 & 41.77 & 0.30 / 0.56 \\
        curl & 1315 & 3119 & 16674 & 824 & 60.9 / 3.52 / 35.56 & 17.11 & 0.55 / 0.26 & 28.40 & 0.43 / 0.34 \\
        dia & 842 & 475 & 3885 & 76 & 48.7 / 7.90 / 43.42 & 31.58 & 0.54 / 0.46 & 47.37 & 0.47 / 0.61 \\
        emacs & 1510 & 5647 & 42073 & 398 & 56.0 / 2.26 / 41.71 & 17.09 & 0.74 / 0.31 & 23.62 & 0.77 / 0.44 \\
        gcc & 83245 & 28510 & 133378 & 1739 & 48.9 / 2.13 / 48.94 & 54.17 & 0.53 / 0.62 & 65.55 & 0.52 / 0.75 \\
        glibc & 20702 & 7042 & 23219 & 325 & 51.69 / 4.62 / 43.69 & 40.00 & 0.62 / 0.56 & 65.54 & 0.52 / 0.78 \\
        gnuplot & 333 & 1447 & 7015 & 126 & 60.3 / 0.79 / 38.89 & 16.67 & 0.57 / 0.24 & 28.57 & 0.56 / 0.40 \\
        hexchat & 358 & 411 & 2089 & 165 & 70.9 / 3.64 / 25.46 & 8.48 & 0.57 / 0.21 & 32.12 & 0.21 / 0.29 \\
        irssi & 504 & 407 & 4287 & 99 & 71.7 / 1.01 / 27.27 & 15.15 & 0.67 / 0.46 & 47.47 & 0.28 / 0.59 \\
        libexpat & 130 & 214 & 1987 & 59 & 59.3 / 5.09 / 35.59 & 15.25 & 1.00 / 0.41 & 38.98 & 0.70 / 0.73 \\
        libsoup & 478 & 321 & 2353 & 159 & 67.3 / 3.77 / 28.93 & 27.67 & 0.43 / 0.58 & 50.31 & 0.33 / 0.79 \\
        libssh-mirror & 349 & 402 & 4644 & 126 & 60.3 / 3.18 / 36.51 & 22.22 & 0.54 / 0.33 & 43.65 & 0.40 / 0.48 \\
        libxml2 & 261 & 2590 & 4522 & 223 & 69.5 / 8.07 / 22.42 & 9.42 & 0.67 / 0.24 & 23.77 & 0.45 / 0.41 \\
        lighttpd1.4 & 282 & 1154 & 4027 & 33 & 45.5 / 9.09 / 45.46 & 24.24 & 0.75 / 0.35 & 33.33 & 0.64 / 0.41 \\
        mapserver & 874 & 1596 & 8787 & 126 & 59.5 / 0.79 / 39.68 & 26.19 & 0.64 / 0.41 & 41.27 & 0.73 / 0.75 \\
        marlin & 7780 & 12429 & 15597 & 1082 & 26.8 / 11.18 / 62.02 & 32.26 & 0.58 / 0.32 & 52.77 & 0.54 / 0.49 \\
        mongo & 70412 & 27905 & 52545 & 686 & 54.1 / 0.73 / 45.19 & 60.20 & 0.33 / 0.60 & 77.26 & 0.36 / 0.84 \\
        openssl & 3507 & 3550 & 22477 & 761 & 63.3 / 5.65 / 31.01 & 18.00 & 0.53 / 0.36 & 36.40 & 0.42 / 0.56 \\
        openvpn & 525 & 994 & 2829 & 138 & 56.5 / 2.90 / 40.58 & 18.12 & 0.76 / 0.33 & 39.13 & 0.52 / 0.48 \\
        ossec-hids & 666 & 823 & 2632 & 88 & 53.4 / 0.00 / 46.59 & 19.32 & 0.76 / 0.37 & 60.23 & 0.42 / 0.63 \\
        retroarch & 7926 & 12053 & 48950 & 536 & 46.3 / 2.61 / 51.12 & 28.36 & 0.70 / 0.39 & 44.59 & 0.65 / 0.56 \\
        totem & 459 & 344 & 4718 & 128 & 73.4 / 2.34 / 24.22 & 29.69 & 0.24 / 0.39 & 45.31 & 0.28 / 0.70 \\
        uwsgi & 301 & 415 & 5189 & 300 & 59.0 / 3.33 / 37.67 & 14.33 & 0.65 / 0.24 & 37.00 & 0.42 / 0.40 \\
        xorg-xserver & 2861 & 3318 & 14826 & 617 & 60.1 / 3.89 / 35.98 & 20.91 & 0.73 / 0.43 & 45.22 & 0.48 / 0.61 \\
        \midrule
        \textbf{Minimum}  & 130 & 214 & 1987 & 33.0 & 26.8 / 0.0 / 22.4 & 8.5 & 0.2 / 0.2 & 23.6 & 0.2 / 0.3 \\
        \textbf{Maximum}  & 83245 & 28510 & 133378 & 1739.0 & 74.3 / 11.2 / 62.0 & 60.2 & 1.0 / 0.6 & 77.2 & 0.8 / 0.8 \\
        \textbf{Average}  & 2261.6 & 3552.3 & 11205.1 & 387.1 & 58.9 / 3.7 / 37.5 & 24.5 & 0.6 / 0.4 & 43.3 & 0.5 / 0.6 \\
        \textbf{Median}  & 525.0 & 407.0 & 4287.0 & 223.0 & 59.5 / 3.2 / 37.7 & 20.9 & 0.6 / 0.4 & 41.8 & 0.5 / 0.6 \\
        \textbf{Mode}  & 126.0 & 60.3 & 1987.0 & 126.0 & 60.3 / 0.8 / - & - & 0.6 / - & - & - / 0.6 \\
        \textbf{Std. Dev.}  & 24648.8 & 9377.7 & 19206.2 & 397.5 & 11.0 / 2.8 / 9.9 & 12.3 & 0.2 / 0.1 & 13.3 & 0.1 / 0.2 \\
        \bottomrule
    \end{tabular}
\end{table*}

\subsection{Data Collection and Evaluation Procedure} \label{sec:collection}

In this section, we explore the essential components of our research methodology, outlining the systematic steps involved in data collection, synthesis, and analysis. The robustness of our study relies on a meticulous process designed to extract meaningful insights from the raw information gathered. %To provide a visual guide to our methodological process, Figure~\ref{fig:Procedure} illustrates a schematic representation of the sequential steps involved in data collection, synthesis, and analysis. 

% \begin{figure}[thb] 
%     \centering
%     \includesvg[width=0.9\linewidth]{Procedure.svg} 
%     \caption{Illustration of data collection and evaluation procedure.}
%     \label{fig:Procedure} 
% \end{figure}

\smallskip

\noindent\textbf{1. Development History Extraction.} We extract the development history from repositories using a custom script developed with pydriller\footnote{\url{https://pydriller.readthedocs.io/en/latest/}}. The script captures essential information, such as file path and the developer responsible for changes, filtering out non-source code files.

\smallskip

\noindent\textbf{2. Variability-Related Information Extraction.} Using pypreprocessor, a Python-based tool, we enhance functionality to identify variable code elements through a line-by-line analysis, excluding files without variable code.

\smallskip

\noindent\textbf{3. Association Between Variability-Related Information and Development History.} We associate developers with variable code components, establishing links between developers, the files they change, and the variabilities they impact.

\smallskip

\noindent\textbf{4. DOA and Ownership Extraction.} We calculate Degree of Authorship (DOA) and Ownership, crafting scripts to process source code files and generate a comprehensive dataset of expertise-related values for each developer in each file considering the entire evolution of the analyzed systems.

%\subsection{Data Evaluation Procedure} \label{sec:eval}

\smallskip

\noindent\textbf{5. Work Specialization (RQ1):} To understand work specialization among developers regarding variable and mandatory code, we used a systematic, longitudinal approach. Monthly, we aggregated a cumulative list of active contributors to identify their involvement with variable code, mandatory code, or both. Developer categories included \textit{Generalists} (exclusively modifying mandatory code), \textit{Specialists} (solely altering variable code components), and \textit{Mixed} developers (engaging with both variable and mandatory code at least once). 

\smallskip

\noindent\textbf{6. Association Between File Expertise and Variable Code (RQ2):} For RQ2, we extracted expertise metrics (e.g., Degree of Authorship, Ownership) for all files. Simultaneously, we compiled a list of developers modifying variable code within each file. Then, we investigated if authors (DOA) and major contributors (Ownership) were among developers changing variable code in a retrospective analysis. Precision and Recall were calculated to assess metric accuracy in identifying relevant developers. Precision measures relevant developers among those recommended, while Recall assesses the proportion of relevant developers correctly identified by the metrics. Precision and Recall gauged the effectiveness of expertise metrics in identifying contributors to variable code, providing insights for improving recommendations.

\section{Preliminary Results} \label{sec:results}

\subsection{Analysis of Work Specialization in Configurable Systems (RQ1)} \label{sec:rq1}

\begin{figure}[thb] 
    \includegraphics[width=\linewidth]{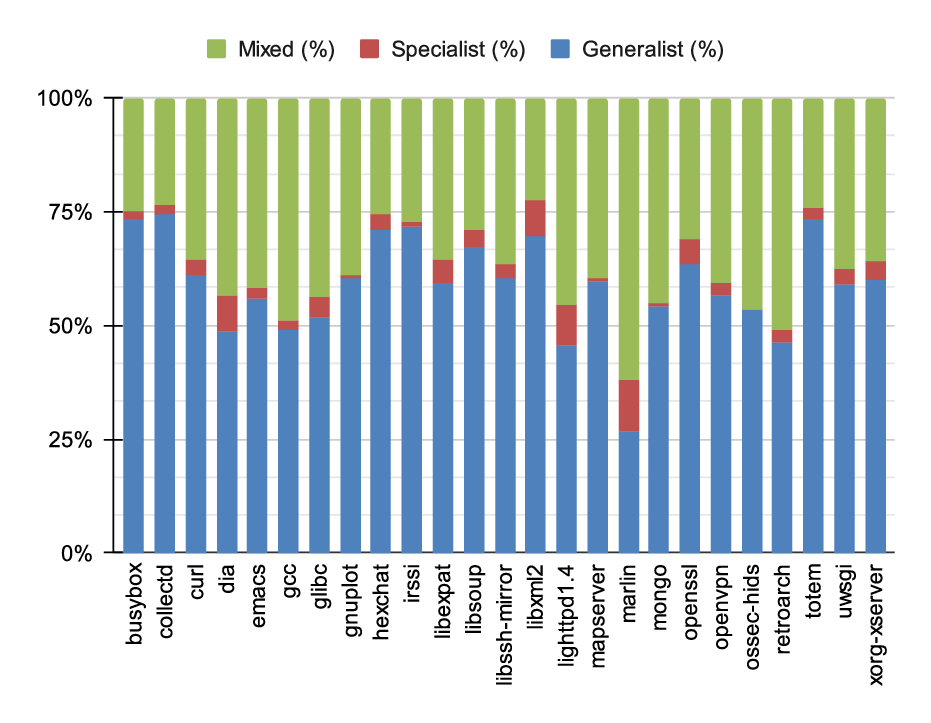} 
    \caption{Work specialization among developers.}
    \label{fig:RQ1} 
\end{figure}

This section investigates work specialization among developers, categorizing them into three profiles: generalists, specialists, and mixed developers based on their involvement in mandatory or variable code. Table~\ref{tab:project-metrics} and Figure~\ref{fig:RQ1} provides an overview, showcasing the total number of developers (\# Devs.) and the corresponding percentages of generalists, specialists, and mixed roles (Generalist / Specialist / Mixed (\%)).

\begin{itemize}
\item \textbf{Generalists:} On average, approximately 59\% of developers fall into this category across subject systems. Their focus on mandatory code may limit exposure to variable components, impacting tasks like maintenance and code reviews involving variable code.

\item \textbf{Specialists:} Accounting for no more than 11.2\% across subject systems, specialists constitute a minority. This suggests a small fraction of developers specializes in variable code development.

\item \textbf{Mixed Developers:} Representing about 37.5\% on average, mixed developers work on both types of code, offering a comprehensive perspective. Their effectiveness varies based on experience and exposure to variable code.
\end{itemize}

In conclusion, understanding the distribution of development efforts in terms of variable and mandatory code is crucial for effective decision-making in CSS. While generalists dominate, their limited exposure to variable code poses challenges. Mixed developers play a pivotal role in maintaining a balance between mandatory and variable code tasks, depending on their experience and knowledge in handling variable code.

\medskip

\begin{tcolorbox}[left=0mm,right=0mm,boxrule=0.1mm,colback=gray!30!white]
\vspace{-0.1cm}
\textit{Lesson Learned 1:} Work specialization among developers of CSS throughout evolution reveals a significant presence of developers primarily focused on mandatory code, potentially limiting their understanding of variable components.
\vspace{-0.1cm}
\end{tcolorbox}

\medskip

\subsection{Association between file expertise and variable code (RQ2)} \label{sec:rq2}

We examined the effectiveness of Degree of Authorship (DOA) and Ownership metrics in identifying developers with expertise in variable code in CSS. Precision denotes the probability of a recommended developer possessing knowledge of variable code. As shown in Table~\ref{tab:project-metrics}, with an average and median precision of 0.6, DOA suggests that the majority of recommended developers are likely to have experience with variable code, reflecting their past involvement in such code. In the case of Ownership, despite a slightly lower precision value (around 0.5), there remains a notable one-in-two chance of identifying a developer with expertise in variable code.

The examination of recall, gauging the proficiency in capturing developers experienced with variable code, exposes consistent challenges. Table~\ref{tab:project-metrics} presents that DOA recall ranges from 0.2 to 0.6, with an average and median of 0.4, revealing a notable trade-off between precision and recall. In comparison, Ownership recall varies between 0.3 and 0.8, showcasing substantial average and median values of 0.6.

The superior recall values in Ownership compared to DOA can be attributed to the ability of the metric to identify a higher number of developers. Table~\ref{tab:project-metrics} further elucidates this difference, highlighting that the average number of developers indicated by Ownership (40\% of the total - Ownership (\%)) exceeds the average indicated by DOA (approximately 25\% of the total - DOA (\%)). Moreover, while Ownership metrics boast higher recall values, this advantage comes at the cost of precision, implying a potential for broader recommendations.

The trade-off between precision and recall in both metrics accentuates the persistent need for refinement in expertise metrics. Striking a balance between high precision and improved recall remains important for effective recommendations and ensuring a more equitable distribution of variable code tasks. %This balance is pivotal to a sustainable approach to the maintenance and evolution of CSS, effectively mitigating the risk of overburdening a small group of developers.

\medskip

\begin{tcolorbox}[left=0mm,right=0mm,boxrule=0.1mm,colback=gray!30!white]
\vspace{-0.1cm}
\textit{Lessons Learned 2:} Expertise metrics may not consistently indicate the expertise of developers, even when variabilities are within the file in which the developer is an expert.
\vspace{-0.1cm}
\end{tcolorbox}

\medskip

\section{Implications} \label{sec:implications}

\noindent \textbf{Striking a balance is paramount: } In the context of CSS, a critical consideration emerges concerning task allocation, particularly for the prevalent group of generalist developers. There is a heightened risk of task accumulation for the limited pool of developers possessing knowledge of variable components. It becomes imperative to not only address potential limitations in the understanding of variable code among generalists but also to mitigate the risk of overburdening the select few who navigate both code types. Exploring strategies for equitable distribution and collaborative knowledge-sharing becomes essential to foster a balanced development environment.

\smallskip

\noindent \textbf{Finding the Expertise Sweet Spot: } The trade-off observed in expertise metrics, emphasizing precision at the expense of recall, underscores the need for refinement in the evaluation of developer expertise. Achieving a balance between high precision, ensuring accurate identification of developers with variable code knowledge, and improved recall, widening the scope of recommendations, is crucial. Refining metrics such as Degree of Authorship (DOA) and Ownership can contribute to more equitable developer recommendations, fostering a system that takes into account the diverse expertise of developers in variable code. Moreover, it is essential to explore and identify the most appropriate thresholds for CSS. The thresholds currently employed in other studies are based on non-CSS, and it may be necessary to adjust them to optimize performance for different systems.
\section{Related Work} \label{sec:relatedwork}

\noindent \textbf{C preprocessor.} 
The C preprocessor has been the subject of substantial criticism in the literature. Numerous studies have discussed the detrimental impact of preprocessor usage on both code quality and maintainability~\cite{ernst2002empirical, kastner2008granularity, medeiros2015love}. Various attempts have been made to extract structural elements from the source code, such as nesting, dependencies, and include hierarchies, and visualize them in dedicated views~\cite{pearse1997experiences, cafeo2016}. Additionally, researchers have explored configuration views that display only portions of feature code~\cite{hofer2010toolchain, kastner2008granularity}. 
%Furthermore, there has been an exploration of innovative ideas like using color-coding to assist developers in working with variable code~\cite{feigenspan2013background}. 
Similar to our work, these studies aim to support the use of the C preprocessor. However, our work focuses on leveraging C preprocessor properties to identify key developers who are likely to excel in maintaining variable code.

\medskip

\noindent \textbf{Expert recommendation.}
In exploring expertise identification and modeling in software development, diverse approaches have been proposed and investigated. McDonald~and~Ackerman Line 10 rule prioritizes developers who last changed a module, echoed by Hossen~et~al. which identifies experts based on recent file changes~\cite{rahman2011ownership, mcdonald2000expertise, hossen2014amalgamating}. 
%Various studies use metrics such as the number of changes to source code elements and investigate expertise in merging operations within development branches~\cite{zanjani2015automatically,ye2021recommending}. Models like Sülün,~Tüzün,~and~Dogrusöz's approach calculate expertise through commits, aiding in code reviewer recommendations~\cite{sulun2019reviewer}. 
Cury~et~al. compared Commit, Blame, and Degree-of-Authorship techniques for identifying file maintainers, while Hannebauer~et~al. assessed algorithms for code reviewer recommendation~\cite{cury2022, hannebauer2016automatically}. Fritz~et~al.'s Degree of Knowledge (DOK) model, incorporating authorship degree and degree of interest, explores knowledge flow within source code history~\cite{fritz2014degree}. Other studies, like Silva~et~al.'s model, consider time-based expertise analysis, while Kagdi~et~al. and Sülün,~Tüzün~and~Dogrusöz incorporate newness in recommending developers for change requests, though not in the context of CSS~\cite{da2015niche}.

\section{Threats to Validity} \label{sec:threats}

 % In terms of \textbf{Conclusion Validity}, issues such as violated assumptions of statistical tests were mitigated by opting for non-parametric tests aligned with data characteristics, while sampling bias in development history extraction was alleviated through the implementation of clear inclusion and exclusion criteria. \textbf{Internal Validity} concerns, including the impact of historical events, were minimized by analyzing systems with many developers, reducing the concentration of knowledge among past reviewers/maintainers, and addressing script and tool limitations through rigorous testing and validation. \textbf{Construct Validity} considerations involved acknowledging and addressing threats like restricted generalizability across constructs and inadequate preoperational explication, with a focus on emphasizing the significance of C language characteristics and providing clear, detailed definitions. \textbf{External Validity} concerns were mitigated by emphasizing global relevance in open-source systems. To enhance generalizability, comprehensive information about subject systems was provided, and real-world, open-source configurable systems on GitHub, extensively studied in prior research, were used.

\textbf{Internal Validity} concerns were minimized by analyzing systems with many developers, reducing the concentration of knowledge among past reviewers/maintainers.
\textbf{Construct Validity} considerations involved acknowledging and addressing threats like restricted generalizability across constructs and inadequate preoperational explication, with a focus on emphasizing the significance of C language characteristics and providing clear, detailed definitions. 
\textbf{External Validity} concerns were mitigated by emphasizing global relevance in open-source systems. To enhance generalizability, comprehensive information about subject systems was provided, and real-world, open-source configurable systems on GitHub, extensively studied in prior research, were used.

\section{Conclusions and Future Work}
\label{sec:conclusion}

Our investigation into work specialization in CSS highlighted the prevalence of generalist developers, raising concerns about their limited exposure to variable code. While specialists play a vital role, their minority status emphasizes the need for a more balanced distribution of expertise. The evaluation of expertise metrics, including Degree of Authorship (DOA) and Ownership, revealed promising precision but highlighted challenges in achieving a balance with recall, necessitating further refinement.

Moving forward, future research should focus on analyzing the concentration of work among mixed and specialist developers. Another goal should be enhancing the knowledge of generalist developers in variable code through targeted interventions. Additionally, exploring different metrics, refining expertise metrics, exploring dynamic task allocation strategies, and developing context-specific evaluation methods tailored to CSS will contribute to a more equitable distribution of development efforts and foster sustainable practices in CSS.

\bibliographystyle{elsarticle-num}
\bibliography{Saner2024CR.bib}

\end{document}